\documentclass[twocolumn,prb]{revtex4}
\usepackage{amsfonts}
\usepackage[T1]{fontenc}
\usepackage{amsmath,amsbsy,amssymb,graphicx}
\usepackage{times}

\begin{document}

\title{ Higher-order topological insulators and semimetals \\
on the breathing Kagome and pyrochlore lattices}
\author{Motohiko Ezawa}
\affiliation{Department of Applied Physics, University of Tokyo, Hongo 7-3-1, 113-8656,
Japan}

\begin{abstract}
A second-order topological insulator in $d$ dimensions is an insulator which
has no $d-1$ dimensional topological boundary states but has $d-2$
dimensional topological boundary states. It is an extended notion of the
conventional topological insulator. Higher-order topological insulators have
been investigated in square and cubic lattices. In this paper, we generalize
them to breathing Kagome and pyrochlore lattices. First, we construct
a second-order topological insulator on the breathing Kagome
lattice. Three topological boundary states emerge at the corner of the
triangle, realizing a 1/3 fractional charge at each corner. Second, we
construct a third-order topological insulator on the
breathing pyrochlore lattice. Four topological boundary states emerge at the
corners of the tetrahedron with a 1/4 fractional charge at each corner.
These higher-order topological insulators are characterized by the quantized
polarization, which constitutes the bulk topological index. Finally, we
study a second-order topological semimetal by stacking the breathing Kagome
lattice.
\end{abstract}

\maketitle

\textit{Introduction:} A topological insulator (TI) in $d$ dimensions has $%
d-1$ dimensional [$(d-1)$D] topological boundary states according to the
bulk-boundary correspondence\cite{Hasan,Qi}. Recently, the concept was
generalized to a higher-order TI (HOTI)\cite%
{Fan,Science,APS,Peng,Lang,Song,Bena,Schin,Liu}. For instance, a
second-order TI is an insulator which has $(d-2)$D topological boundary
states but no $(d-1)$D topological boundary states. Namely, the boundary of
the second-order TI is an ordinary TI. Similarly, a third-order TI is an
insulator which has $(d-3)$D boundary states but no $(d-1)$D and $(d-2)$D
boundary states. It implies that the boundary of the third-order TI is the
second-order TI. The HOTI is characterized by the bulk topological index\cite%
{Science,Bena,Song}. It belongs to a special class of topological insulators
to which the conventional bulk-boundary correspondence is not applicable. It
is intriguing that there are several HOTIs previously considered to be
trivial insulators. So far, HOTIs have been studied for the square and cubic
lattices\cite{Science,APS,Peng,Lang,Song,Bena,Schin}.

\begin{figure}[t]
\centerline{\includegraphics[width=0.49\textwidth]{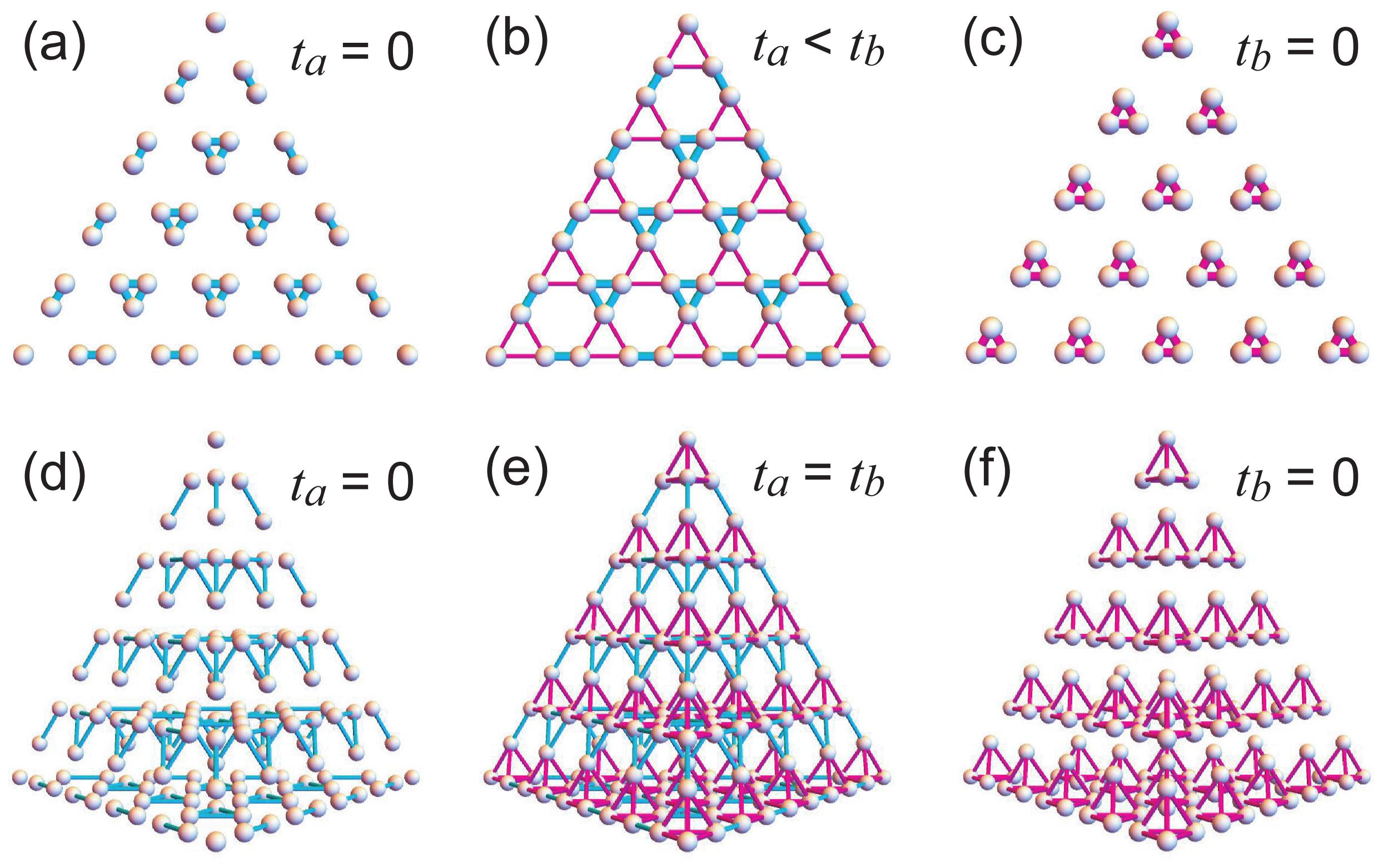}}
\caption{Illustration of triangles made of the breathing Kagome lattice with
(a) $t_{a}=0$, (b) $0<t_{a}<t_{b}$ and (c) $t_{b}=0$. A triangle contains
many small triangles. There are three isolated atoms at the corner of the
triangle for $t_{a}=0$, while there are no isolated atoms for $t_{b}=0$. The
size of the triangle is $L=5$. Illustration of tetrahedrons made of the
breathing pyrochlore lattice with (d) $t_{a}=0$, (e) $t_{a}=t_{b}$ and (f) $%
t_{b}=0$. A tetrahedron contains many small tetrahedrons. There are four
isolated atoms at the corner of the tetrahedron for $t_{a}=0$, while there
are none for $t_{b}=0$. The size of the tetradedron is $L=5$. See Fig.%
\protect\ref{FigFCC}(b) for the unit cell of the pyrochlore lattice ($t_a=t_b
$). }
\label{FigIllust}
\end{figure}

In this paper, we propose HOTIs on the breathing Kagome lattice and the
breathing pyrochlore lattices. They have attracted much attention in the
context of the spin system\cite{Shca,Udagawa,Repe,Owe,Bz,Ben,Li,Sava,Tsune},
and experimentally been realized\cite%
{Va,Her,Hiroi,Kimura,Hiroi2,Hiroi3,Hiroi4,Rau,Haku2,Tanaka,Matsuda}. The
structures of the breathing Kagome and pyrochlore lattices are illustrated
in Fig.\ref{FigIllust}. First, with respect to the breathing Kagome lattice
[Fig.\ref{FigIllust}(b)], we find no topological boundary states in the 1D
geometry (i.e., nanoribbon) but find three topological boundary states in
the 0D geometry (i.e., triangle). A 1/3 fractional charge emerges when we
put one electron into the zero-energy states. Second, with respect to the
breathing pyrochlore lattice [Fig.\ref{FigIllust}(e)], we find no
topological boundary states in the 2D and 1D geometries but find four
topological boundary states in the 0D geometry (i.e., tetrahedron). A 1/4
fractional charge emerges when we put one electron into the zero-energy
states. Finally, we construct a second-order topological semimetal by
stacking the breathing Kagome lattice.

In the present models the bulk topological index is a polarization, which is
the integral of the Berry connection, as in the previous models\cite%
{Bena,Science,Song}. When the C$_3$ and mirror symmetries are intact, it is
quantized and distinguishes the trivial and topological phases.
Equivalently, the mismatch between the Wannier center and the lattice site
distinguishes the trivial and topological phases. Zero-energy corner modes
emerge in the topological phase.

\begin{figure*}[t]
\centerline{\includegraphics[width=0.98\textwidth]{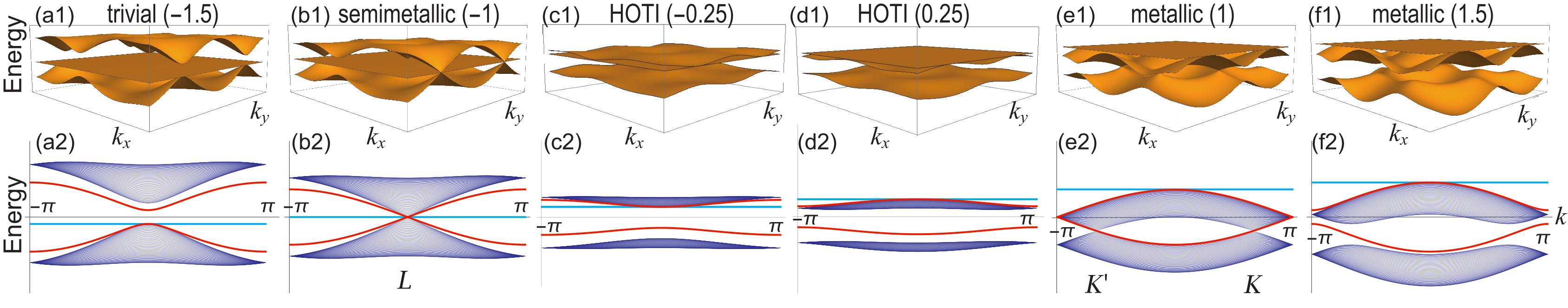}}
\caption{Band structure of breathing Kagome lattices with (a1) $%
t_{a}/t_{b}=-1.5$, (b1) $t_{a}/t_{b}=-1$, (c1) $t_{a}/t_{b}=-0.25$, (d1) $%
t_{a}/t_{b}=0.25$, (e1) $t_{a}/t_{b}=1$ and (f1) $t_{a}/t_{b}=1.5$. The
horizontal axes are $k_{x}$ and $k_{y}$. (a2)--(f2) The corresponding band
structure of nanoribbons, where the horizontal axis is $k$. In each figure,
two red curves represent boundary modes while a cyan line represents a
perfect flat band belonging to the bulk. Although the nanoribbon spectrum
indicates that the bulk must be a trivial insulator for (c1) and (d1), it is
actually a HOTI. }
\label{KagomeRibbon}
\end{figure*}

\textit{Second-order TIs in the breathing Kagome lattice: } We consider a
fermion model on the breathing Kagome lattice. The bulk Hamiltonian is given
by 
\begin{equation}
H=-\left( 
\begin{array}{ccc}
0 & h_{12} & h_{13} \\ 
h_{12}^{\ast } & 0 & h_{23} \\ 
h_{13}^{\ast } & h_{23}^{\ast } & 0%
\end{array}
\right) ,  \label{H3}
\end{equation}
with $h_{12}=t_{a}+t_{b}e^{-i\left( k_{x}/2+\sqrt{3}k_{y}/2\right) }$, $%
h_{13}=t_{a}+t_{b}e^{-ik_{x}}$ and $h_{23}=t_{a}+t_{b}e^{i\left( -k_{x}/2+%
\sqrt{3}k_{y}/2\right) }$, where we have introduced two hopping parameters $%
t_{a}$ and $t_{b}$ corresponding to upward and downward triangles, as shown
in Fig.\ref{FigIllust}. We have taken a gauge such that the $t_{a}$-terms
contain no phase factor in the unit cell. It is set $t_{b}>0$ without loss
of generality. We analyze the system by making four-step arguments.

(i) First, we examine the bulk band spectrum. The dispersion relation reads%
\cite{Udagawa} 
\begin{equation}
E=t_{a}+t_{b},-\frac{t_{a}+t_{b}}{2}\pm \frac{1}{2}\sqrt{9\left(
t_{a}^{2}+t_{b}^{2}\right) -6t_{a}t_{b}+8t_{a}t_{b}F},
\end{equation}
with $F=\cos k_{x}+2\cos \frac{k_{x}}{2}\cos \frac{\sqrt{3}}{2}k_{y}$. There
is a flat band at $E=t_{a}+t_{b}$. The band gap closes at the $K=(2\pi /3,0)$
and $K^{\prime }=(-2\pi /3,0)$ points for $t_{a}=t_{b}$ [See Fig.\ref%
{KagomeRibbon}(e1)] and at the $\Gamma =(0,0)$ point for $t_{a}=-t_{b}$ [See
Fig.\ref{KagomeRibbon}(b1)]. It is an insulator for $t_{a}/t_{b}<-1$\ and $%
-1<t_{a}/t_{b}<1/2$, while it is metallic for $t_{a}/t_{b}>1/2$.

\begin{figure}[t]
\centerline{\includegraphics[width=0.44\textwidth]{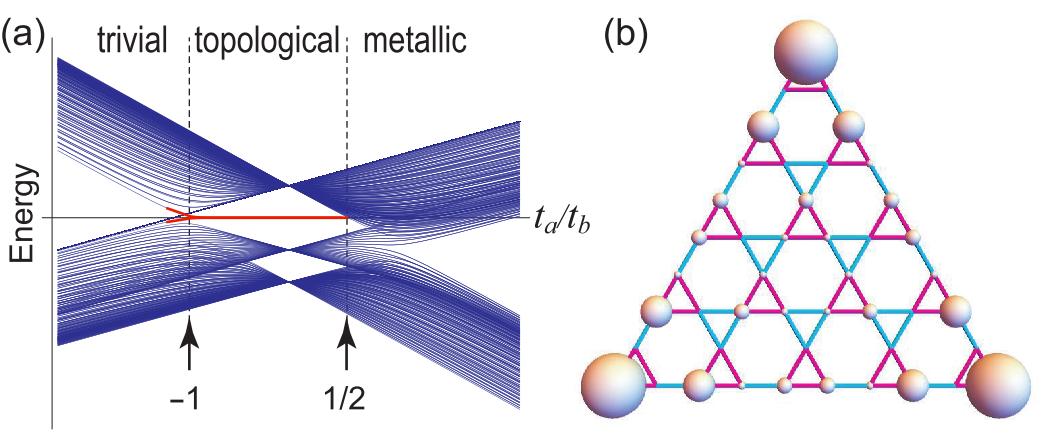}}
\caption{(a) Energy spectrum of the triangle made of the breathing Kagome
lattice with $L=20$. The horizontal axis is $t_{a}/t_{b}$. There emerge
zero-energy states (marked in red) for $-1<t_{a}/t_{b}<1/2$. They are
topological boundary states. (b) The square root of the local density of
states $\protect\sqrt{\protect\rho _{i}}$ for the triangle with $L=5$ and $%
t_{a}/t_{b}=1/2$. The amplitude is represented by the radius of the spheres.
The local density of states becomes arbitrarily small except for the three
corners for $L\gg 1$. }
\label{FigTri}
\end{figure}

(ii) Second, we investigate nanoribbons\cite{Nanoribbon} made of the
breathing Kagome lattice, which corresponds to the $(d-1)$D geometry with $%
d=2$. We show the band structure in Fig.\ref{KagomeRibbon} for typical
values of $t_{a}/t_{b}$. According to the conventional bulk-boundary
correspondence, the bulk must be a trivial insulator\cite{Phos} both for $%
t_{a}/t_{b}<-1$\ and $-1<t_{a}/t_{b}<1/2$. However, this is not the case, as
we now show.

(iii) Third, we investigate nanodisks\cite{Nanodisk} made of the breathing
Kagome lattice, which corresponds to the $(d-2)$D geometry with $d=2$. As a
nanodisk respecting the C$_3$ and mirror symmetries, we consider a triangle
containing many small triangles whose directions are opposite, as
illustrated in Fig.\ref{FigIllust}(a)--(c). We define the size $L$ of the
triangle by the number of small triangles along one edge at $t_{b}=0$: See
Fig.\ref{FigIllust}(c). By using this definition, there are $2L $ atoms at
the boundary and $3L(L+1)/2$ atoms in total. We show the energy spectrum as
a function of $t_{a}/t_{b}$ for $-1.5<t_{a}/t_{b}<1.5$ for $L=20$ in Fig.\ref%
{FigTri}. The energy spectrum changes smoothly by changing $t_{a}/t_{b}$
smoothly. Zero-energy states emerge for $-1<t_{a}/t_{b}<1/2$. We show the
square root of the local density of states of the zero-energy states in Fig.%
\ref{FigTri}(b). It is well localized at the three corners of the triangle.
The phenomenon is similar to the case of the square lattice with dimerized
hoppings\cite{Science}, where four corner atoms are isolated.

The system is exactly solvable at $t_{a}=0$ and $t_{b}=0$, where we may
calculate the energy spectrum analytically. On one hand, when $t_{a}=0$,
there are three isolated atoms, $L-1$ dimers and $(L-1)(L-2)/2$ trimers.
Isolated atoms have the zero-energy, the dimers have the energy $\pm t_{a}$
and the trimers have the energy $-2t_{a}$ and two fold $t_{a}$. As a result,
there are three zero-energy states, $L(L-1)/2$ energy levels with $E=t_{a}$
and $(L-1)(L-2)/2$ energy levels with $E=-t_{a}$. On the other hand, when $%
t_{b}=0$, there are $L(L+1)/2$ trimers and there are no isolated atoms and
dimers, implying that the zero-energy states do not appear. Our numerical
analysis shows that the zero-energy states emerge for $-1<t_{a}/t_{b}<1/2$.

We now argue that these zero-energy states are topological boundary states.
We focus on the system at $t_{a}=0$. In this case the boundary of the
breathing Kagome lattice is detached completely from the bulk and forms a 1D
dimerized chain: See Fig.\ref{FigIllust}(a). It is described by the
Su-Schrieffer-Heager (SSH) model. Since the SSH model describes a TI, we
conclude that the boundary of the breathing Kagome lattice is a TI at $%
t_{a}=0$. Since the energy spectrum at $t_{a}\neq 0$ is adiabatically
connected to that at $t_{a}=0$, the zero-energy states are topological
boundary states even for $t_{a}\neq 0$.

As shown in Fig.\ref{FigTri}(b), the local density of states is separated
equally in the three corners. When we put one electron into the zero-energy
states, the $1/3$ fractional charge appears at the three corner of the
triangle. This is an extension of the 1/2 fractional charge in the SSH model
describing a Polyacetylene\cite{JR}.

(iv) Finally, we show the existence of the bulk topological index
characterizing the HOTI. In the case of the SSH model, the polarization $%
p_{x}$ along the $x$ axis is the bulk topological index, which is protected
by the mirror symmetry along the $x$ direction. We generalize it into higher
dimension\cite{Science,Bena,Song}.

There are three mirror symmetries for the breathing Kagome lattice. They are
the mirror symmetries $M_{x}$ with respect to the $x$ axis, and $M_{\pm }$
with respect to the two lines obtained by rotating the $x$ axis by $\pm 2\pi
/3$. The polarization along the $x_{i}$ axis is the expectation value of the
position, 
\begin{equation}
p_{i}=\frac{1}{S}\int_{\text{BZ}}A_{i}d^{2}\mathbf{k},  \label{PolarP}
\end{equation}%
where $A_{i}=-i\left\langle \psi \right\vert \partial _{k_{i}}\left\vert
\psi \right\rangle $ is the Berry connection with $x_{i}=x,y$ and $S=8\pi
^{2}/\sqrt{3}$ is the area of the Brillouin zone. The set of the
polarization ($p_{x},p_{y}$) is identical to the Wannier center, which is
the expectation value of the Wannier function obtained by making the Fourier
transformation of the Bloch function\cite{Bena,Science,Song}. Note that $%
p_{x}$\ is defined mod 1 according to the formula (\ref{PolarP}) since the
it changes by an integer under gauge transformation. We similarly obtain the
polarization $p_{\pm }=-p_{x}/2\pm \sqrt{3}p_{y}/2$ along the other lines.

By taking into account the C$_3$ and mirror symmetries, we analyze the
quantity 
\begin{equation}
P_{3}=p_{x}^{2}+p_{+}^{2}+p_{-}^{2}=\frac{3}{2}\left(
p_{x}^{2}+p_{y}^{2}\right) ,
\end{equation}%
which measures the distance of the Wannier center from the origin. It is
protected by the three mirror symmetries. As we shall soon see, we evaluate
it as $P_{3}=0$ in the insulator phase for $t_{a}/t_{b}<-1$ 
while $P_{3}=1/2 $ in the insulator phase for $-1<t_{a}/t_{b}<1/2$. Consequently,
the system on the breathing Kagome lattice is a second-order TI for $-1<t_{a}/t_{b}<1/2$, 
where $P_{3}$ is the bulk topological index.

\begin{figure}[t]
\centerline{\includegraphics[width=0.49\textwidth]{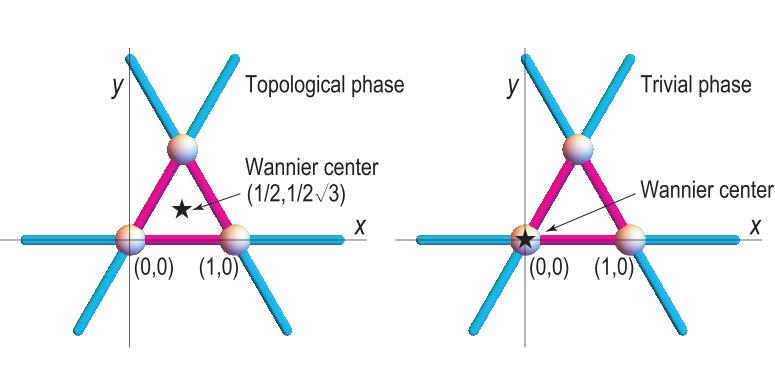}}
\caption{The Wannier centers in the topological and trivial phases. }
\label{FigWannier}
\end{figure}

\begin{figure}[t]
\centerline{\includegraphics[width=0.44\textwidth]{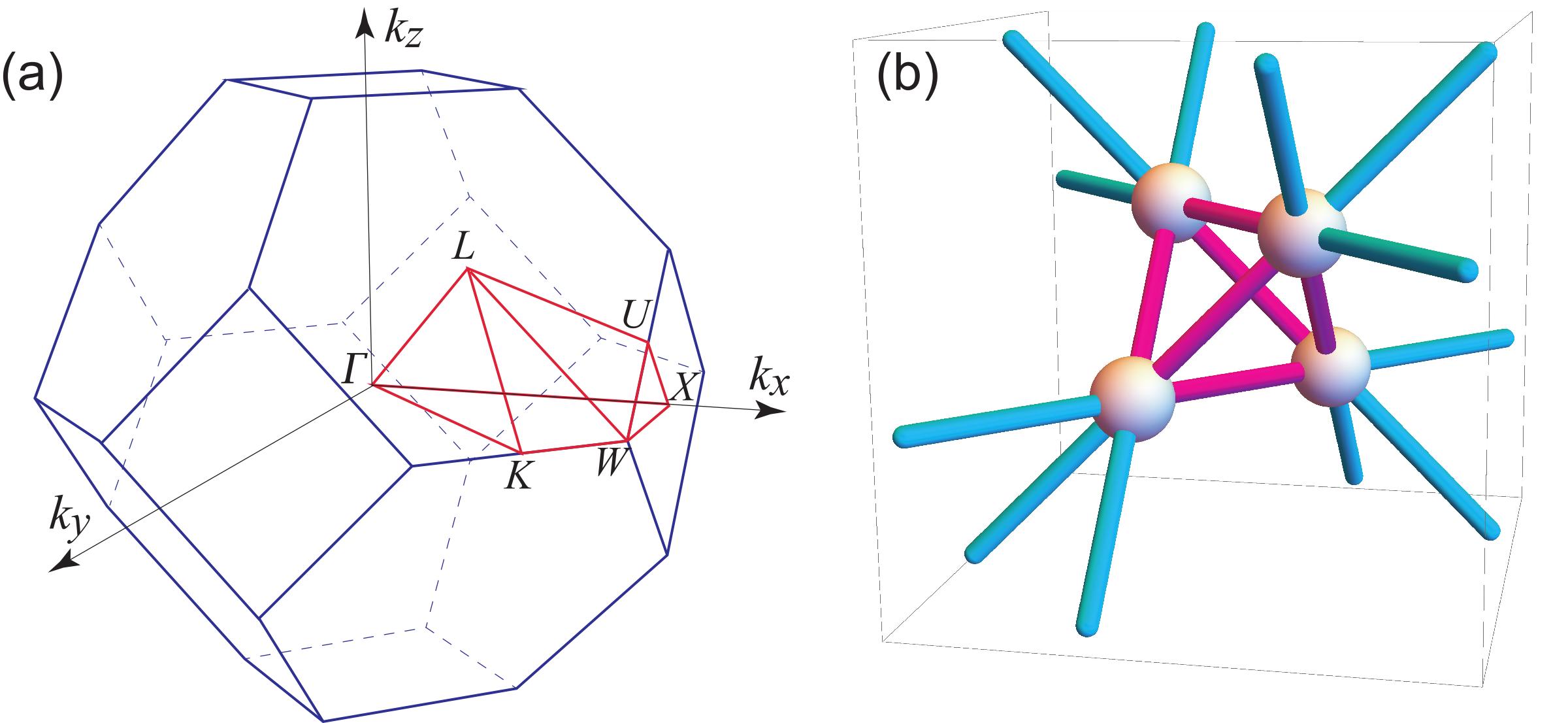}}
\caption{(a) Brillouin zone of the pyrochlore lattice. Letters $\Gamma $, $X$, 
$W$, $U$, $L$ and $K$ represent high symmetry points. (b) Unit cell of the
pyrochlore lattice ($t_{a}=t_{b}$).}
\label{FigFCC}
\end{figure}

Due to the mirror symmetry $M_{x}$ it follows that $p_{x}=-p_{x}$. 
Since $p_{x}$ is defined mod $1$, we solve this as $p_{x}=0$ or $1/2$. Similarly we
find that $p_{\pm }=0$ or $1/2$. Hence, the bulk topological index $P_{3}$
is quantized. It cannot change its value unless the gap closes.
Consequently, it takes a constant value in each topological phase. It is
enough to calculate $P_{3}$ at any one point in the topological or trivial
phase.

First, we set $t_{a}=0$ to calculate $P_{3}$ for the topological phase ($-1<t_{a}/t_{b}<1/2$). 
The ground state wave function is found to be $\psi
=\left( 1,e^{i\left( k_{x}+\sqrt{3}k_{y}\right) /2},e^{ik_{x}}\right) ^{t}/\sqrt{3}$, 
with which we calculate the Berry connection as $A_{x}=1/2$ 
and $A_{y}=1/2\sqrt{3}$. It follows that $p_{x}=1/2$, $p_{+}=0$, $p_{-}=1/2$ and 
$P_{3}=1/2$. We note that the relation $p_{x}=p_{+}=p_{-}$ does not hold
since the C$_{3}$\ symmetry is broken by the choice of the coordinate: See
Fig.\ref{FigWannier}. The Wannier center exists at the center 
$(p_{x},p_{y})= $ $(1/2,1/2\sqrt{3})$ of the small triangle. The mismatch
between it and the lattice site produces the zero-energy boundary states at
the corner in the topological phase\cite{Bena,Science,Song}.

Next, by choosing $t_{b}=0$ for the trivial phase, the wave function is
given by $\psi =\left( 1,1,1\right) ^{t}/\sqrt{3}$, and we find $A_{x}=0$
and $A_{y}=0$, from which it follows that $P_{3}=0$. Consequently the
Wannier center is on the lattice site $(0,0)$: See Fig.\ref{FigWannier}.

\begin{figure*}[t]
\centerline{\includegraphics[width=0.98\textwidth]{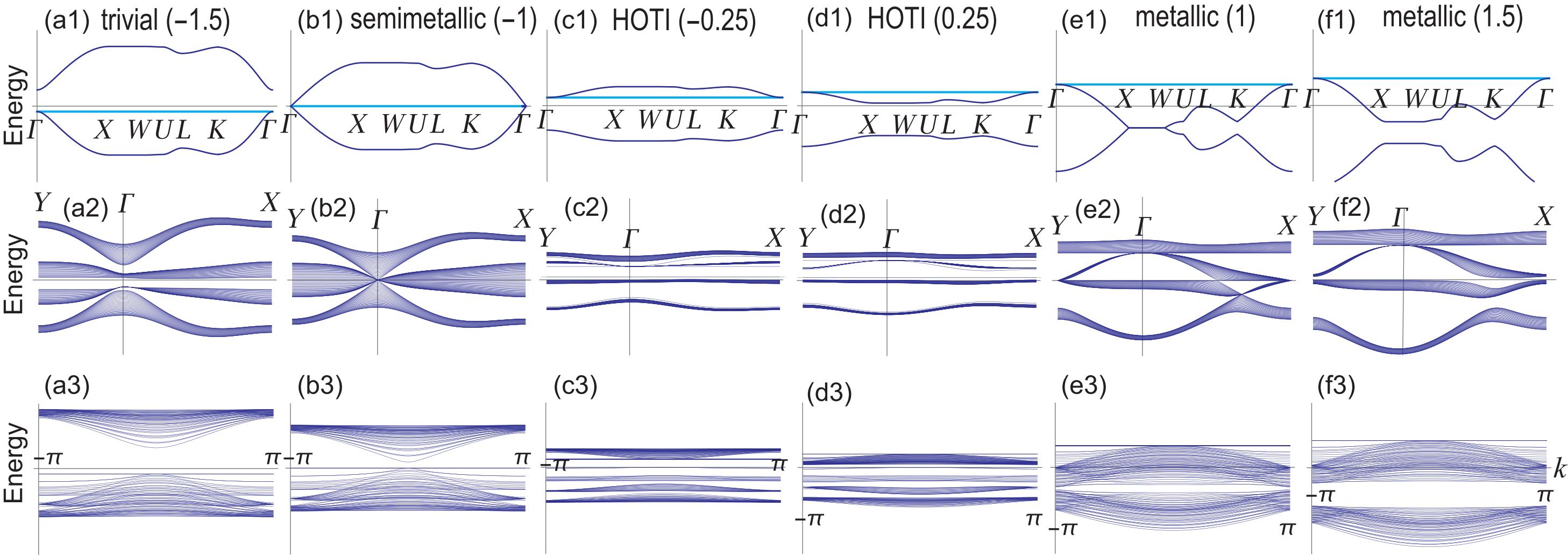}}
\caption{Band structure of breathing pyrochlore lattices along the 
$\Gamma $-$X$-$W$-$U$-$L$-$K$-$\Gamma $ line. (a1) $t_{a}/t_{b}=-1.5$, 
(b1) $t_{a}/t_{b}=-1$, (c1) $t_{a}/t_{b}=-0.25$, (d1) $t_{a}/t_{b}=0.25$, 
(e1) $t_{a}/t_{b}=1$ and (f1) $t_{a}/t_{b}=1.5$. Cyan lines represent perfect flat
bands belonging to the bulk. (a2)--(f2) The corresponding band structure of
a thin film, where the horizontal axis is momentum along the $X$-$\Gamma $-$Y $ line. 
(a3)--(f3) The corresponding band structure of a triangular prism,
whose size of the triangle is $L=9$. There are no topological boundary
states for thin films and triangular prisms. Nevertheless, it is not a
trivial insulator but a HOTI for (c1) and (d1).}
\label{PyroBand}
\end{figure*}

\begin{figure}[t]
\centerline{\includegraphics[width=0.49\textwidth]{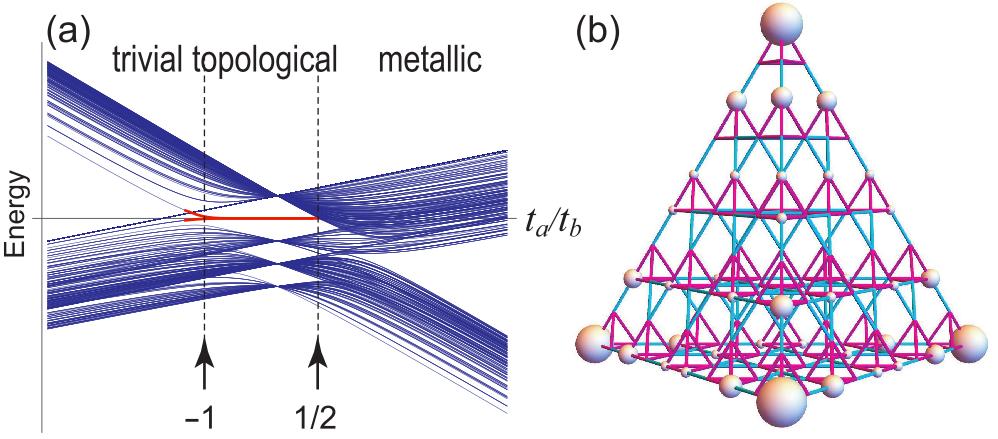}}
\caption{(a) Energy spectrum of the tetrahedron made of the breathing
pyrochlore lattice with $L=9$. The horizontal axis is $t_{a}/t_{b}$. There
emerge zero-energy states (marked in red) for $-1<t_{a}/t_{b}<1/2$. They are
topological boundary states. (b) The square root of the local density of
states $\protect\sqrt{\protect\rho _{i}}$ for the tetrahedron with $L=5$ and 
$t_{a}/t_{b}=1/2$. The amplitude is represented by the radius of the
spheres. The local density of states becomes arbitrarily small except for
the four corners for $L\gg 1$. The localized states emerge only at the four
corners of the tetrahedron. }
\label{FigTetra}
\end{figure}

\textit{Third-order TIs in the breathing pyrochlore lattice:} We proceed to
investigate the third-order TI. A natural extension of the breathing Kagome
lattice into three dimensions is the breathing pyrochlore lattice [Fig.\ref{FigIllust}(d)--(f)]. 
The Brillouin zone and the unit cell of the pyrochlore
lattice are shown in Fig.\ref{FigFCC}. The Hamiltonian is given by 
\begin{equation}
H=-\left( 
\begin{array}{cccc}
0 & h_{12} & h_{13} & h_{14} \\ 
h_{12}^{\ast } & 0 & h_{23} & h_{24} \\ 
h_{13}^{\ast } & h_{23}^{\ast } & 0 & h_{34} \\ 
h_{14}^{\ast } & h_{24}^{\ast } & h_{34}^{\ast } & 0%
\end{array}
\right)
\end{equation}
with $h_{12}=t_{a}+t_{b}e^{-i\left( k_{x}+k_{y}\right) /2}$, $%
h_{13}=t_{a}+t_{b}e^{-i\left( k_{y}+k_{z}\right) /2}$, $%
h_{14}=t_{a}+t_{b}e^{-i\left( k_{z}+k_{x}\right) /2}$, $%
h_{23}=t_{a}+t_{b}e^{-i\left( k_{z}-k_{x}\right) /2}$, $%
h_{24}=t_{a}+t_{b}e^{-i\left( -k_{y}+k_{z}\right) /24}$ and $%
h_{34}=t_{a}+t_{b}e^{-i\left( k_{x}-k_{y}\right) /2}$. We analyze the system
by making four-step arguments.

(i) The bulk spectrum is given by the two perfect flat bands\cite{Udagawa}, 
$E=t_{a}+t_{b}$, and 
\begin{equation}
E=-t_{a}-t_{b}\pm \sqrt{t_{a}^{2}+t_{b}^{2}-t_{a}t_{b}+t_{a}t_{b}G},
\end{equation}
with $G=\cos \frac{k_{x}}{2}\cos \frac{k_{y}}{2}+\cos \frac{k_{y}}{2}\cos 
\frac{k_{z}}{2}+\cos \frac{k_{z}}{2}\cos \frac{k_{x}}{2}$. The band
structure is shown along the $\Gamma $-$X$-$W$-$U$-$L$-$K$-$\Gamma $ line in
Fig.\ref{PyroBand}(a1)--(f1) for typical values of $t_{a}/t_{b}$. The band
gap closes at $t_{a}/t_{b}=\pm 1$. It is an insulator for $t_{a}/t_{b}<-1$ 
and $-1<t_{a}/t_{b}<1/2$, while it is metallic for $t_{a}/t_{b}>1/2$.

(ii) We investigate thin films and triangular prisms made of the breathing
pyrochlore lattice, which correspond to the $(d-1)$D and $(d-2)$D geometries
with $d=3$. We show their band structures in Fig.\ref{PyroBand} (a2)--(f2)
and (a3)--(f3) for typical values of $t_{a}/t_{b}$. According to the
conventional bulk-boundary correspondence, the bulk must be a trivial
insulator both for $t_{a}/t_{b}<-1$\ and $-1<t_{a}/t_{b}<1/2$. However, this
is not the case, as we now show.

(iii) Next, we investigate tetrahedrons shown in Fig.\ref{FigIllust}(e),
which correspond to the $(d-3)$D geometry. A tetrahedron contains many small
tetrahedrons whose directions are different, as illustrated in Fig.\ref{FigIllust}(d)--(f). 
It has four triangular faces made of the breathing
Kagome lattice. We define the size $L$ of the tetrahedron by the number of
small tetrahedrons along one edge at $t_{b}=0$: See Fig.\ref{FigIllust}(f).
There are $2L(L+1)(L+2)/3$ atoms in total. The energy spectrum is given for 
$L=9$ in Fig.\ref{FigTetra}(a). As in the case of the breathing Kagome
lattice, zero-energy states emerge for $-1<t_{a}/t_{b}<1/2$. We show the
square root of the local density of states of the zero-energy states in 
Fig.\ref{FigTetra}(b). It is localized at the four corners of the tetrahedron,
where there emerge 1/4 fractional charges.

We note that the system is exactly solvable for the special cases of 
$t_{a}=0 $ and $t_{b}=0$. One one hand, for $t_{a}=0$, there are four
isolated atoms at the corners of the tetrahedron, which contribute to the
four zero-energy states. Additionally, there are $3(L-1)$ dimers, 
$3(L-1)(L-2)/2$ trimers and $L(L-1)(L-2)/6$ tetramers. On the other hand, for 
$t_{b}=0$, there are $L(L+1)(L+2)/6$ tetramers, which lead to no zero-energy
states. Our numerical analysis shows that the zero-energy states emerge for $-1<t_{a}/t_{b}<1/2$.

We argue that these zero-energy states are topological boundary states. When 
$t_{a}=0$, the 2D boundary is detached completely from the bulk and each
face forms a triangle made of the breathing Kagome lattice: See Fig.\ref{FigIllust}(d). 
We have already shown that it possesses topological
zero-energy states localized at the corners. Since the energy spectrum 
at $t_{a}\neq 0$ is adiabatically connected to that at $t_{a}=0$, the
zero-energy states are topological boundary state even for $t_{a}\neq 0$.
Namely, the zero-energy states found in Fig.\ref{FigTetra} are attributed to
the zero-energy states in Fig.\ref{FigTri}.

(iv) Finally, we discuss the bulk topological index. As in the case of the
breathing Kagome lattice, since there are six mirror symmetries, $M_{x+y}$, 
$M_{y+z}$, $M_{z+x}$, $M_{x-y}$, $M_{y-z}$ and$\;M_{z-x}$, we analyze the
quantity 
\begin{eqnarray}
P_{6}
&=&p_{x+y}^{2}+p_{y+z}^{2}+p_{z+x}^{2}+p_{x-y}^{2}+p_{y-z}^{2}+p_{z-x}^{2} 
\notag \\
&=&4\left( p_{x}^{2}+p_{y}^{2}+p_{z}^{2}\right) .
\end{eqnarray}
We now show that it is given by $P_{6}=3$ for the topological phase and 
$P_{6}=0$ for the trivial phase. The breathing pyrochlore lattice is a
third-order TI for $-1<t_{a}/t_{b}<1/2$, where $P_{6}$ is the bulk
topological index.

In the topological phase, the wave function for $t_{a}=0$ is given by $\psi
=\left( 1,e^{-i\left( k_{x}+k_{y}\right) },e^{-i\left( k_{y}+k_{z}\right)
},e^{-i\left( k_{z}+k_{x}\right) }\right) ^{t}/2$, from which we obtain the
Berry connection as $A_{x}=A_{y}=A_{z}=1/2$ and $P_{6}=3$. The Wannier
center exists at the center $(1/2,1/2,1/2)$ of the small tetrahedron.

On the other hand, in the trivial phase, the wave function for $t_{b}=0$ is
given by $\psi =\left( -1,1,1,1\right) ^{t}/2$, from which we obtain the
Berry connection as $A_{x}=A_{y}=A_{z}=0$ and $P_{6}=0$. The Wannier center
exists at the lattice site $(0,0,0)$.

\textit{Second-order topological semimetals:} Recently a second-order
topological semimetal is constructed by stacking square lattices with
dimerized hoppings\cite{Liu}. It is a 3D bulk semimetal where the gap closes
linearly at two points. An interesting feature is that there is no Fermi arc
in the $(d-1)$D geometry with $d=3$ (i.e., thin film with 2D boundaries).
However, 1D zero-energy boundary states connecting the two gap closing
points appear in the $(d-2)$D geometry (i.e., square prism).

In the similar way, we can construct a second-order topological semimetal by
stacking the breathing Kagome lattice. By replacing $t_{a}\mapsto
t_{a}+t_{z}\cos k_{z}$ in the Hamiltonian (\ref{H3}), we obtain the
Hamiltonian for the second-order topological semimetal. The band gap closes
at $t_{a}+t_{z}\cos k_{z}=-t_{b}$. There is no zero-energy state for the
boundary, while there are zero-energy states for the boundary of the
boundary for $k_{z}$ satisfying $-1<(t_{a}+t_{z}\cos k_{z})/t_{b}<1/2$.

We have shown that the breathing Kagome and pyrochlore lattices are HOTIs
together with the emergence of zero-energy corner modes. 
However, they may shift in energy by a local potential at the corners due to the lack of the chiral symmetry in the models. 
Nevertheless, the emergence of the corner modes are protected by the C$_{3}$ and mirror symmetries.

The author is very much grateful to N. Nagaosa for helpful discussions on
the subject. This work is supported by the Grants-in-Aid for Scientific
Research from MEXT KAKENHI (Grant Nos.JP17K05490 and JP15H05854). This work
is also supported by CREST, JST (JPMJCR16F1).

\end{document}